\documentclass[aps,twocolumn,prl,groupedaddress,showpacs]{revtex4-1}

\usepackage{hyperref}
\usepackage{graphicx}

\begin{document}


\title{Quantum error correction in a solid-state hybrid spin register.} 

\author{G. Waldherr$^1$$^{\dagger}$}\email{g.waldherr@physik.uni-stuttgart.de}
\author{Y. Wang$^1$$^{\dagger}$}
\author{S. Zaiser$^1$}
\author{M. Jamali$^1$}
\author{T. Schulte-Herbr\"uggen$^2$}
\author{H. Abe$^3$}
\author{T. Ohshima$^3$}
\author{J. Isoya$^4$}
\author{P. Neumann$^1$}
\author{J. Wrachtrup$^1$}

\affiliation{$^1$3. Physikalisches Institut, Research Center SCOPE, and MPI for Solid State Research, University of Stuttgart, Pfaffenwaldring 57, 70569 Stuttgart, Germany}
\affiliation{$^2$Department of Chemistry, Technical University of Munich (TUM), D-85747 Garching, Germany}
\affiliation{$^3$Japan Atomic Energy Agency, Takasaki, Gunma 370-1292, Japan}
\affiliation{$^4$Research Center for Knowledge Communities, University of Tsukuba, Tsukuba, Ibaraki, 305-8550 Japan}
\affiliation{$^{\dagger}$These authors contributed equally to this work.}

\date{\today}

\begin{abstract}
Hybrid quantum systems seek to combine the strength of its constituents to master the fundamental conflicting requirements of quantum technology: fast and accurate systems control together with perfect shielding from the environment, including the measurements apparatus, to achieve long quantum coherence.
Excellent examples for hybrid quantum systems are heterogeneous spin systems where electron spins are used for readout and control \cite{kane_silicon-based_1998, morton_solid-state_2008, neumann_single-shot_2010, koehl_room_2011, pla_single-atom_2012, taminiau_detection_2012, pfaff_demonstration_2012, dolde_room-temperature_2013, chekhovich_nuclear_2013, bernien_heralded_2013} while nuclear spins are used as long-lived quantum bits \cite{maurer_room-temperature_2012,pla_high-fidelity_2013,yin_optical_2013}. 
Here we show that joint initialization, projective readout and fast local and non-local gate operations are no longer conflicting requirements in those systems, even under ambient conditions.
We demonstrate high-fidelity initialization of a whole spin register ($99\%$) and single-shot readout of multiple individual nuclear spins by using the ancillary electron spin of a nitrogen-vacancy defect in diamond. 
Implementation of a novel non-local gate generic to our hybrid electron-nuclear quantum register allows to prepare entangled states of three nuclear spins, with fidelities exceeding $85\%$.
An important tool for scalable quantum computation is quantum error correction \cite{shor_fault-tolerant_1996, cory_experimental_1998, knill_benchmarking_2001, boulant_experimental_2005, moussa_demonstration_2011, schindler_experimental_2011, reed_realization_2012}.
Combining, for the first time, optimal-control based error avoidance with error correction, we realize a three-qubit phase-flip error correction algorithm. 
Utilizing optimal control, all of the above algorithms achieve fidelities approaching fault tolerant quantum operation, thus paving the way to large scale integrations. 
Our techniques can be used to improve scaling of quantum networks relying on diamond spins \cite{dolde_room-temperature_2013,bernien_heralded_2013}, phosphorous in silicon \cite{pla_high-fidelity_2013} or other spin systems like quantum dots \cite{le_gall_optical_2011}, silicon carbide \cite{koehl_room_2011} or rare earth ions in solids \cite{yin_optical_2013}.
\end{abstract}

\maketitle

For increasing size of a quantum system, decoherence due to environmental noise becomes especially severe:
As the number of possible coherences increases exponentially with the number of qubits, so does the rate of decoherence.
Errors induced by decoherence cannot be avoided, as even small changes of a quantum state lead to finite error probabilities, contrary to classical computing, where small state changes (e.g. voltage) do not affect the final result.
Thus, even for extremely isolated systems like nuclear spins, scalable quantum information processing (QIP) imposes a virtually impossible endeavour.
A solution for scalable QIP is quantum error correction (QEC).
The threshold theorem states that the final error probability of a quantum algorithm on a large register can be made arbitrarily small, if the probability of an error on a single qubit is below a certain threshold, and QEC is applied \cite{shor_fault-tolerant_1996}. 
This makes QEC a necessary requirement for scalable quantum information technology.

As a model system for the implementation of QEC in a hybrid spin system we use the nitrogen-vacancy defect (NV) in diamond \cite{gruber_scanning_1997, dutt_quantum_2007, neumann_multipartite_2008, neumann_single-shot_2010, robledo_high-fidelity_2011, pfaff_demonstration_2012, taminiau_detection_2012, maurer_room-temperature_2012, dolde_room-temperature_2013, bernien_heralded_2013}. 
Nuclear spins in the carbon lattice are addressed, read out and initialized by the electron spin of the defect, which can be polarized and readout by optical means.

\begin{figure*}
\includegraphics[width=0.98\textwidth]{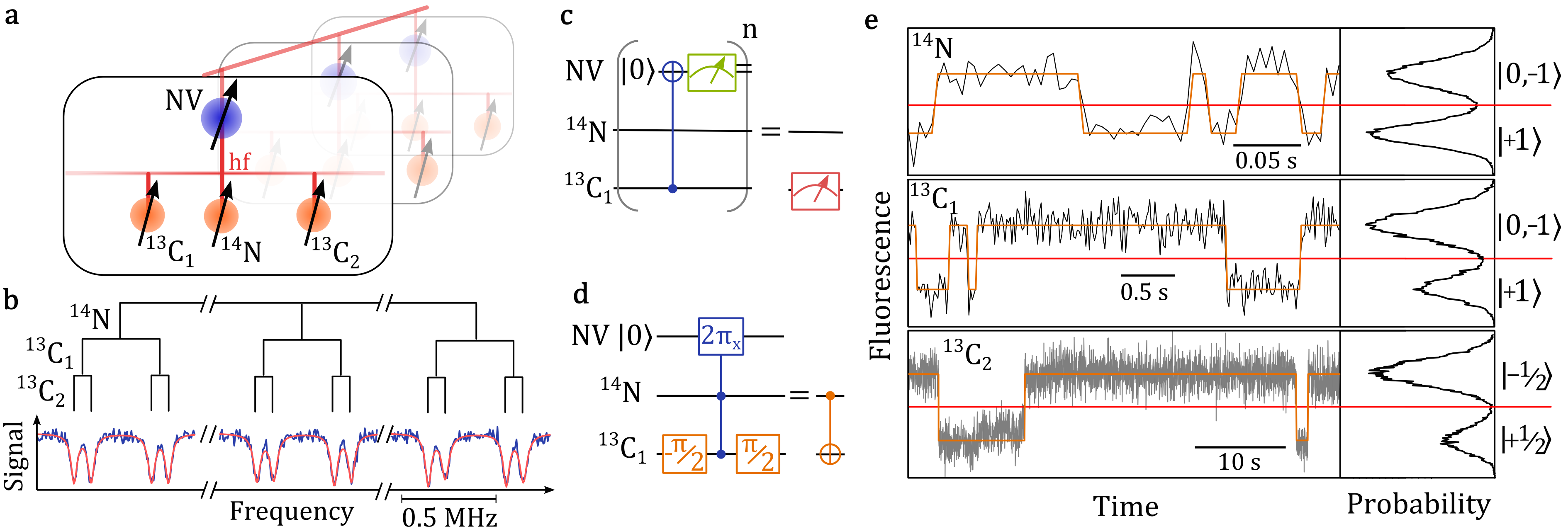}
	\caption{\textbf{Single shot readout and control of a nuclear spin register via the NV.}
	\textbf{a}, Illustration of the quantum register consisting of three nuclear spins hyperfine (hf) coupled to the central NV electron spin, enabling nuclear spin readout and entanglement.
	The system is a prototype single node of a larger register, coupled by the electron spins \cite{dolde_room-temperature_2013,bernien_heralded_2013}.
	\textbf{b}, Hyperfine splitting of the $m_S = 0$ to $-1$ transition into 12 individual lines ($^{14}$N: 2.16 MHz, $^{13}$C$_1$: 413 kHz,  $^{13}$C$_2$: 89 kHz).
	\textbf{c}, Wire diagram for the projective readout of the nuclear spins, here exemplified for a $^{13}$C spin.
	First, the state of one nuclear spin is mapped onto the electron spin, by initialization of the electron spin and a C$_{\rm{n}}$NOT$_{\rm{e}}$ gate, as described in the methods section.
	This correlates the fluorescence from the NV with the nuclear spin state.
	While the electron spin state is destroyed (polarized) during the readout, this can be prevented for the nuclear spin, enabling quantum non-demolition measurements \cite{neumann_single-shot_2010}.
	By repeating this readout sequence, projective, non-destructive measurements of the nuclear spin can be achieved.
	\textbf{d}, Wire diagram of a CNOT gate between two nuclear spins via the NV electron spin.
	Here, a dot corresponds to the $\left|1\right\rangle$ state.
	\textbf{e}, Left side: Time trace of the NV fluorescence during repetitive readout of the corresponding nuclear spin as shown in c.
	The jumps indicate flips of the nuclear spin, and show that the readout of all nuclear spins is much faster than the flip rate, which allows for projective readout.	
	Right side: Histograms of measurement results of the time traces and threshold for single-shot readout.
	For initialization, the threshold can be shifted down to increase the initialization fidelity, at the expense of the rate of successful initialization events.
	}
\label{fig:1}
\end{figure*} 
Specifically, we use three nuclear spins ($^{14}$N, $I=1$ and two $^{13}$C, $I=\frac{1}{2}$) as quantum bits which are hyperfine (hf) coupled to the electron spin (Fig. \ref{fig:1}a), within an isotopically purified diamond with $\approx 0.2\%$ $^{13}$C.
All nuclear spins of the register can be individually addressed via the hf interaction with the electron spin, which shows sprectral signatures of the 12 states of the nuclear register (Fig. \ref{fig:1}b).
Projective, non-destructive readout of the nuclear spins requires strong magnetic fields (here 6200 Gauss) and suitable electron nuclear spin coupling parameters. 
Projective, single-shot readout is achieved by correlating the electron spin readout signal with the nuclear qubit state by a state specific C$_{\rm{n}}$NOT$_{\rm{e}}$ gate.
The corresponding measurement sequence is shown in Fig. \ref{fig:1}c.
As seen from Fig. \ref{fig:1}e, the lifetime of the nuclear spins' states is much longer than the time needed to measure the state. 
To further quantify the fidelity of our readout, Fig. \ref{fig:1}e also shows histograms of measurement results for nuclear spin readout.
By placing a threshold between the two distributions, the state of the spin is determined by measuring whether the photon count is below or above this threshold.
Here, we achieve readout fidelities of $^{14}$N: $95.8\%$, $^{13}$C$_1$: $96.9\%$, $^{13}$C$_2$: $99.6\%$.

Next, we create three-qubit entangled states of the nuclear spins.
Projective readout can be used for high-fidelity initialization of the register.
However, the success rate of measurement based initialization scales inversely with the number of register states.
To overcome this limitation, we first apply polarization transfer with the electron spin \cite{dutt_quantum_2007}.
With a combination of these two methods (Fig. \ref{fig:2}a), we achieve an initialization fidelity of the whole register up to $99\%$ (see Supplementary Information).
Additionally, initialization of the NV into its negative charge state is required, which is implemented via charge state post-selection (see Supplementary Information). 
\begin{figure*}
\includegraphics[width=0.95\textwidth]{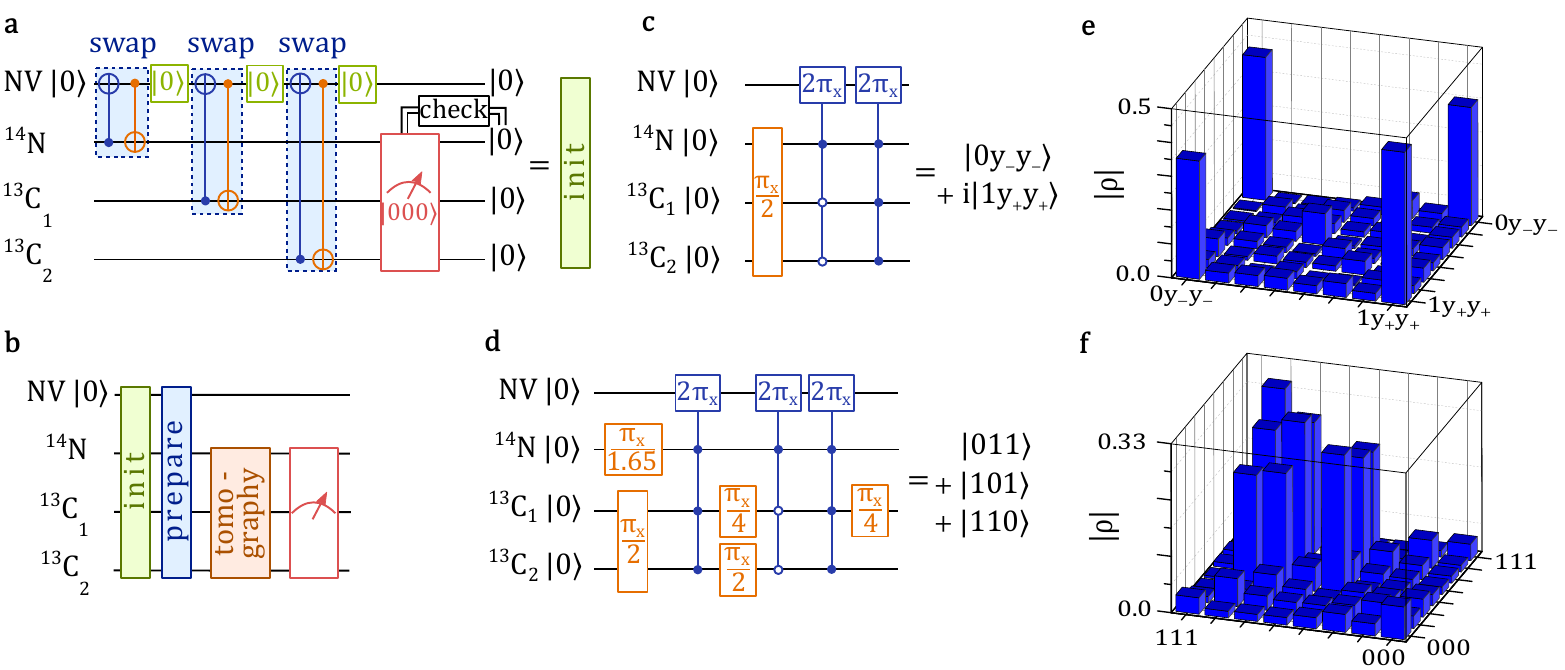}
	\caption{\textbf{Preparation and tomography of three-qubit entangled states}
	\textbf{a}, Initialization sequence for the nuclear spin register.
	The NV electron spin can be initialized non-unitarily by applying green laser illumination, which is indicated by the $|0\rangle$ gate.
	Swap gates are used to transfer electron spin polarization onto the nuclear spins.
	A final measurement probes whether the state of the nuclear spins is $|000\rangle$.
	\textbf{b}, Measurement sequence.
	The tomography step consists of local gates on the nuclear spins.
	The preparation step is either sequence c) for the GHZ state or sequence d) for the W state.
	\textbf{c}, Reconstructed density matrices for the GHZ state and f) for the W state.
	Only absolute values are shown.}
\label{fig:2}
\end{figure*}

Previously reported entanglement of two nuclear spins relied on pseudo-initialization \cite{neumann_multipartite_2008}, or non-deterministic entanglement-by-measurement \cite{pfaff_demonstration_2012} at cryogenic temperatures. 
Instead, we use a novel scheme to create entanglement using interactions with the electron spin.
The scheme is based on a controlled 2$\pi_x$ rotation of the electron spin conditional on the state of the nuclear register, which effectively acts as a CPhase gate.
Thereby, the phase of the latter state is shifted by $\pi$ (e.g. $|00\rangle \rightarrow -|00\rangle$), allowing for nuclear CNOT gates (Fig. \ref{fig:1}d, an example for the creation of an entangled state is shown in the methods section).
Utilizing this gate, we create a GHZ-like entangled state (Figs \ref{fig:2}c,e) and a W-state (Figs \ref{fig:2}d,f) of the 3 nuclear spins.
The advantage of our approach is that it avoids the fast decohering subspace of the electron spin, while still exploiting hf interaction. 
Utilizing the interaction between electron and nuclear spins instead of direct dipole-dipole interaction between nuclei yields ultra-fast control of nuclear spins compared to the standard nuclear magnetic resonance control method.
It is thus essential to high fidelity gate operations in electron nuclear hybrid spin quantum systems.
To achieve robust, high-fidelity operation of the CPhase gate we apply optimal control \cite{machnes_comparing_2011}. 
Optimal control offers a general framework for creating robust state manipulation, which can employ both geometric phase gates \cite{wu_geometric_2013} (offering intrinsic robustness) and dynamically corrected gates \cite{khodjasteh_dynamically_2009}.
In our case, main sources of gate errors are frequency detuning and crosstalk. 

After preparation, the state of the nuclear spin register is measured by state tomography (see Supplementary Information).
The resulting density matrices for the GHZ and W states are shown in Figs \ref{fig:2}e,f.
The estimated fidelity $F = \left\langle\psi\left|\rho\right|\psi\right\rangle$ of the final state $\rho$ with ideal state $\left|\psi\right\rangle$ before the measurement is $F = (88\pm1)\%$ for the GHZ state and $F=(85\pm3)\%$ for the W state, corrected by the readout fidelity of $89\%$.
Factors reducing the fidelity apart from the actual entanglement sequence are $T_1$ decay of the electron spin ($\approx 96\%$) and initialization and post-selection fidelity ($\approx 97\%$).

To show genuine three-qubit entanglement, we also measure violation of the Mermin inequality \cite{mermin_extreme_1990}.
For the GHZ-like state in Fig. \ref{fig:2}c, the inequality is $\left|\left\langle\sigma_x\sigma_x\sigma_z\right\rangle+\left\langle\sigma_x\sigma_z\sigma_x\right\rangle+\left\langle\sigma_y\sigma_x\sigma_x\right\rangle-\left\langle\sigma_y\sigma_z\sigma_z\right\rangle\right| \leq 2.$ (cf. Supplementary Information), which holds for any local, realistic theory.
This inequality is violated by quantum mechanics, yielding a value of maximal 4.
For the present case, we measured a value $3.258 \pm 0.014$.

To demonstrate the functionality of our hybrid quantum register we perform quantum error correction with the three nuclear spins.
Despite being a corner stone for scalable quantum systems, QEC has only been implemented on one solid state systems, namely superconducting circuits \cite{reed_realization_2012}.
This is due to challenging demands of high-fidelity non-local control of at least three qubits.
Generally, errors can occur in the form of bit flips, which act as $\left|0\right\rangle \leftrightarrow \left|1\right\rangle$, or phase flips, which act as $\left|x_+\right\rangle \leftrightarrow \left|x_-\right\rangle$, $\left|y_+\right\rangle \leftrightarrow \left|y_-\right\rangle$.
Apparently, phase flips are identical to bit flips, however in a different basis of the qubit.
The error correction protocol for bit flips and phase flips thus is essentially identical, only differing by local $\pi_x/2$ rotations to change the basis in which the state is stored. 
QEC is based on encoding the information of one physical qubit into two states of a multi-qubit register, forming one logical qubit.
The number of errors which can be corrected is given by the bit flip distance $d$ of the two logical states.
As an example, for three qubits and the logical states $\left|000\right\rangle$ and $\left|111\right\rangle$, three bit flips ($d=3$) are necessary to convert one state into the other.
Thus, erroneous states resulting from single bit flips ($d=1$) stay always closer to the original state and can therefore be corrected by a majority vote principle, whereas multiple flips ($d>1$) will inevitably destroy the information.
While errors in quantum mechanics appear as continuous state rotations, the subsequent error detection projects these rotations onto the two possible cases of not having an error or having an error on the qubit, with probability given by the angle of rotation.
Thus, after the detection, possible errors can be reverted to restore the original state.

\begin{figure}
\includegraphics[width=0.48\textwidth]{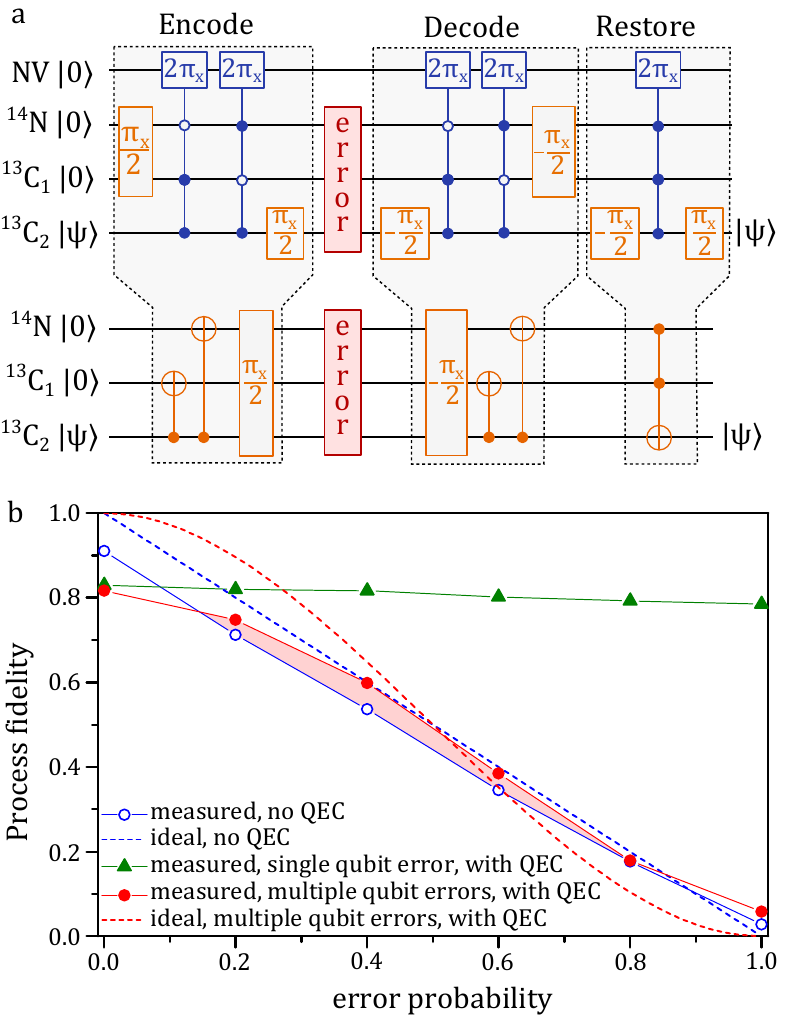}
	\caption{\textbf{Quantum error correction protocol and measurement.}
	\textbf{a}, Quantum circuit for phase flip error correction.
	Upper part: Gates as implemented in the experiment, lower part: Effective gates on the nuclear spins (cf. Fig. \ref{fig:1}d).
	Controlled errors on a qubit are introduced by changing the phase of all operations on that qubit after the occurrence of the error.
	For conditional gates, a dot corresponds to the $\left|1\right\rangle$ state and a circle to the $\left|0\right\rangle$ state.
	\textbf{b}, Measured process fidelity $f$ of the QEC protocol (see SI), in dependence of the error probability $p$.
	For reference, the blue circles and the dashed blue line are the process fidelity without error correction, for our measurements and for ideal measurements ($f(p) = 1 - p$), respectively.
	For the measurement, the total duration of the sequence was the same as for the QEC, for better comparison with the effective performance of error correction.
	To demonstrate that our QEC code prevents single qubit errors, errors on only one qubit (here $^{13}$C$_2$) were induced (green triangles).
	Furthermore, we apply QEC while introducing errors on all qubits, with the same probability of having an error for each qubit.
	The red dots show the measured process fidelity, and the dashed red line shows the theoretical ideal case with $f(p) = 1 - 3p^2 + 2p^3$.
	}
\label{fig:3}
\end{figure}

Here, we demonstrate a phase flip correction protocol, since the rate of phase flip errors is usually much higher than the rate of bit flip errors.
The QEC sequence is shown in Fig. \ref{fig:3}a.
This code does not need error detection measurements:
It corrects and restores the state of one qubit, and transfers any errors on this qubit onto the two other qubits by coherent evolution, as shown below.
The original state $\left|\psi\right\rangle = \alpha\left|0\right\rangle+\beta\left|1\right\rangle$ is encoded as $\alpha\left|y_+y_+y_+\right\rangle + \beta\left|y_-y_-y_-\right\rangle$.
After decoding, the state will be $\alpha\left|000\right\rangle + \beta\left|111\right\rangle$ in case of no error, and $\alpha\left|001\right\rangle + \beta\left|110\right\rangle$ in case of an error on e.g. $^{13}$C$_2$.
Then, the error is detected by the other two qubits, by flipping them conditional on the state of $^{13}$C$_2$, yielding $\left|00\right\rangle\otimes\left(\alpha\left|0\right\rangle + \beta\left|1\right\rangle\right)$ and $\left|11\right\rangle\otimes\left(\alpha\left|1\right\rangle + \beta\left|0\right\rangle\right)$, respectively.
Finally, the state of $^{13}$C$_2$ is flipped conditional on the state of the two other spins, resulting in $\left|00\right\rangle\otimes\left(\alpha\left|0\right\rangle + \beta\left|1\right\rangle\right)$ if there was no error, or $\left|11\right\rangle\otimes\left(\alpha\left|0\right\rangle + \beta\left|1\right\rangle\right)$ if there was an error, which is, in both cases, the original state $\left|\psi\right\rangle$ for $^{13}$C$_2$.

To obtain the performance of our implementation, we measure the fidelity of the quantum process on the main qubit (cf. Supplementary Information), depending on the probability that an error occurs.
Fig. \ref{fig:3}b shows the results of our measurements.
Indeed, single qubit errors are corrected by the algorithm (green triangles).
The added complexity of the correction reduces the fidelity for very low error probability.
However, even for simultaneous errors on all qubits (red dots), the application of error correction surpasses the case where no correction is applied (blue circles), if the error probability increases (red area in Fig. \ref{fig:3}b).

In conclusion, our experiments demonstrate a fully functional hybrid quantum register.
We have combined optimal-control based error avoidance with error correction, where in particular, error correction based on multiple nuclear spins will greatly facilitate scaling of such systems.
While we exemplified fast and high fidelity entanglement gates on diamond nuclear spins, our method is applicable to other hybrid spin quantum system for example in silicon \cite{pla_high-fidelity_2013}, silicon carbide \cite{koehl_room_2011}, quantum dots \cite{le_gall_optical_2011} or rare-earth ions in solids \cite{yin_optical_2013}.
For strongly coupled nuclear spins (hf interaction $>$ inhomogeneous broadening) as used here, more than five such qubits are usable per single electron spin (see Supporting Information) where five qubits is the requirement for full quantum error correction.
Including weakly coupled nuclear spins (hf interaction $>$ homogeneous broadening) can even increase the size of the register by a factor of three or four \cite{zhao_sensing_2012, taminiau_detection_2012}.
Most importantly, utilizing nuclear spins with very weak hf interaction allows for repetitive error correction because quantum information on the latter spins is preserved while the remaining register is reset \cite{maurer_room-temperature_2012}.

\section{Methods}
\footnotesize
\textbf{Experimental setup.}
The measurements were performed with a home-built confocal microscope.
A 532 nm DPSS laser is focused via an oil objective onto the NV to excite its electron.
Fluorescence is filtered by a 650 nm long-pass filter and a 50 $\mu$m pinhole, and then detected by a single photon counting avalanche photo diode.
The sample is a type IIa chemical vapour deposition diamond.
NVs are formed by electron irradiation and subsequent annealing.
The magnetic field is created by a permanent magnet attached to a 3-d positioning system, to align the magnetic field parallel to the NV axis.
RF and MW signals are generated by an arbitrary waveform generator.
Additionally, MW signals are mixed with a high frequency source to obtain high frequency arbitrary waveforms at around 15 GHz.
Signals are applied via a coplanar waveguide microstructure, which is build directly onto the diamond nearby the NV, such that optical access to the NV is possible through the gap of the waveguide.
\\
\\
\textbf{Non-local gates.}
Non-local gates are applied via the electron spin, which can be initialized by laser illumination \cite{gruber_scanning_1997}.
As can be seen in Fig. \ref{fig:1}b, the MW transition frequency of the electron spin is distinct for the 12 possible states of the nuclear register.
By applying frequency-selective MW pulses, operations on the electron spin can be carried out conditional on any state (or any combination of states) of the register.
A C$_{\rm{n}}$NOT$_{\rm{e}}$ is implemented by a $\pi$ rotation of the electron spin, conditioned on the desired nuclear spin states.
Here, the $\pi$ rotation commences around the $x$-axis of the Bloch sphere if not otherwise indicated.
\\
In the following, we illustrate the mechanism on the creation of a two qubit GHZ-like state using a nuclear CPhase gate based on a conditional 2$\pi_x$ rotation of the electron spin.
We start with both nuclear qubits initialized in state $|00\rangle$, and apply $\pi_x$/2 rotations on both, to end up in (excluding normalization)
\begin{equation}\label{eq:1}
|00\rangle - i|01\rangle - |10\rangle + i|11\rangle.
\end{equation}
Now, a CPhase conditional on the nuclear spin state being $|11\rangle$ will change the phase of this state to $-|11\rangle$.
This yields the entangled GHZ-like state
\begin{equation}\label{eq:2}
|00\rangle - i|01\rangle - |10\rangle - i|11\rangle = |0Y_-\rangle + |1Y_+\rangle,
\end{equation}
where $Y_{-/+} = |0\rangle \mp |1\rangle$.
This method allows the implementation of nuclear CNOT gates (Fig. \ref{fig:1}d).
\\
\\
\textbf{Note} After submission of this manuscript we became aware of related work by Taminiau et. al., \href{http://lanl.arxiv.org/abs/1309.5452}{arXiv:1309.5452} in which 3-qubit quantum error correction is implemented with the NV using weakly coupled nuclear spins.


\textbf{Acknowledgments} We thank F. Dolde, I. Jakobi, M. Kleinmann, F. Jelezko, J. Honert, A. Brunner for experimental help and fruitful discussions.
We acknowledge financial support by the ERC project SQUTEC, the DFG SFB/TR21, the EU projects DIAMANT, SIQS and QESSENCE, JST-DFG (FOR1482), as well as the Volkswagenstiftung.


\end{document}



\title{Supplementary Information: \\
Quantum error correction in a solid-state hybrid spin register}

\author{G. Waldherr$^1$$^{\dagger}$}\email{g.waldherr@physik.uni-stuttgart.de}
\author{Y. Wang$^1$$^{\dagger}$}
\author{S. Zaiser$^1$}
\author{M. Jamali$^1$}
\author{T. Schulte-Herbr\"uggen$^2$}
\author{H. Abe$^3$}
\author{T. Ohshima$^3$}
\author{J. Isoya$^4$}
\author{P. Neumann$^1$}
\author{J. Wrachtrup$^1$}

\affiliation{$^1$3. Physikalisches Institut, Research Center SCOPE, and MPI for Solid State Research, University of Stuttgart, Pfaffenwaldring 57, 70569 Stuttgart, Germany}
\affiliation{$^2$Department of Chemistry, Technical University of Munich (TUM), D-85747 Garching, Germany}
\affiliation{$^3$Japan Atomic Energy Agency, Takasaki, Gunma 370-1292, Japan}
\affiliation{$^4$Research Center for Knowledge Communities, University of Tsukuba, Tsukuba, Ibaraki, 305-8550 Japan}
\affiliation{$^{\dagger}$These authors contributed equally to this work.}

%

\date{\today}

\pacs{}
\maketitle

\begin{figure*}
\includegraphics[width=0.85\textwidth]{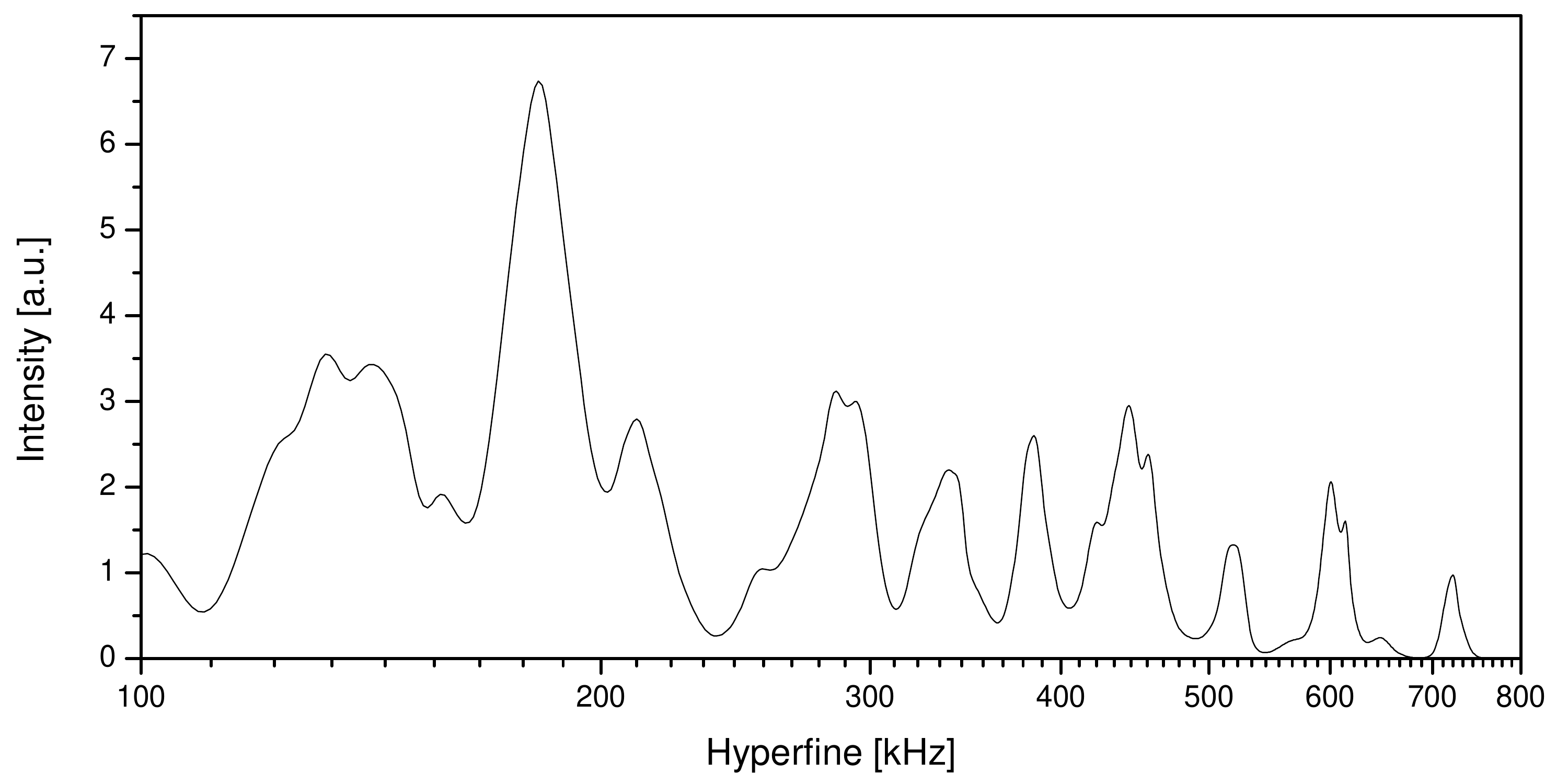}
	\caption{Hyperfine spectrum as explained in the text.
	Only hyperfine values with an error of less than 4\% were used, such that hyperfine values which are close-by can be resolved.
	}
\label{fig:hfspectrum}
\end{figure*}
\section{$^{13}$C hyperfine interactions}
To find an NV with two $^{13}$C nuclear spins suitable for single shot readout (SSR), we acquired statistics of possible $^{13}$C hyperfine splittings for almost 3300 NVs.
For each analysed NV, the EPR spectrum is measured, as shown in fig. 1b of the main manuscript. 
A multi-peak function fit was used to determine hyperfine interactions larger than approximately 100 kHz, as the line width of the transition is around 50 kHz. 
These measurements where performed at magnetic fields of around 10 Gauss.
However, the magnetic field was not aligned for all NVs, which can slightly alter the measured hyperfine interaction compared to the value at zero field or at small, aligned magnetic fields.
The deviation is expected to be below  2.5 \% (in the worst case of pure dipolar hyperfine interaction, unfavorable $^{13}$C position and perpendicular magnetic field).
Finally, we can create a hyperfine probability spectrum in the following way:
For each observed hyperfine interaction, we take a Gaussian distribution at the centre of the observed value, and a standard deviation according to the fit accuracy.
Summing up over all measured values results in the spectrum shown in Fig. \ref{fig:hfspectrum}.
By trying to perform single shot readout on these $^{13}$C spins, we found that the hyperfine splitting (in kHz) of 124, 211, 384, 422, 517 are usable for single shot readout at magnetic fields of around 0.5 T, with spin lifetimes (in readout steps) comparable to or better than $^{14}$N \cite{neumann_single-shot_2010}.

\section{Estimate of strongly and weakly coupled $^{13}$C spins}
In the main text a fast and high fidelity CPhase gate is presented, which is used for example for entanglement generation.
This gate relies on the individual addressability of all conditional electron spin transitions, for example transitions $m_S=0 \leftrightarrow -1$.
This means conditional on all possible nuclear spin states $\left| m_{I,1}, m_{I,2}, \ldots, m_{I,N} \right\rangle$ of all $N$ coupled nuclear spins, where the $m_{I,i}$ are the nuclear spin projections).
For $N$ nuclei there are $2^N$ such transitions.
The inhomogeneous linewidth of the NV electron spin transition can be small enough to resolve hyperfine interactions on the order of a few kHz \cite{maurer_room-temperature_2012}.
We call nuclei which are stronger coupled than the inhomogeneous linewidth strongly coupled spins.
The maximum suitable hyperfine interaction of the closest $^{13}$C spins is a few MHz.
Therefore, in principle up to $\sim 10^3$ resonance lines are resolvable, resulting in $\approx 10$ addressable nuclear spins.
Thus, the presented high speed and high fidelity CPhase gate use suitable for a maximum number of coupled nuclear spins of around 10.
The above estimation, however, assumes that each of the mentioned coupled nuclear spins, if ordered by their coupling strength, increase their coupling strength by roughly a factor of two with respect to the previous nucleus.
Although these coupling strengths are in principle available \cite{dreau_single-shot_2013}, the formation of such a register would almost require the deliberate construction of the diamond surrounding the NV center atom by atom.
However, the situation is much less involved if a little bit less nuclei are required, due to the exponential dependence on $N$.
For example full quantum error correction, which requires five nuclei \cite{knill_benchmarking_2001}, is accessible because the number of the above mentioned electron spin transitions in that case is no more than 32 (a factor of $\approx 3$ more than for the present register).

The number of usable nuclear spins can be increased even further via addressing in the nuclear spin resonance frequency space and by including weakly coupled nuclear spins.
In that case the number of transitions scales much more conveniently (i.e. linearly with $N$).
However, then the presented fast CPhase gate is not practicable anymore. 
For detection and manipulation of weakly coupled $^{13}$C nuclear spins with the electron spin via dynamical decoupling \cite{zhao_sensing_2012}, the limiting factor is the homogeneous decoherence time $T_2$ of the electron spin.
The $T_2$ time depends linearly on the concentration $c$ of $^{13}$C \cite{mizuochi_coherence_2009}.
Therefore, the maximum distance $r_{\rm{max}}$ of a detectable $^{13}$C to the NV is $1/r_{\rm{max}}^3 \propto T_2 \propto c$, due to the magnetic dipole-dipole hyperfine interaction.
The average number $N$ of $^{13}$C nuclei within the radius $r_{\rm{max}}$ is $N \propto c r_{\rm{max}}^3$, such that $N$ is independent of the $^{13}$C concentration.
Note that for very low $^{13}$C concentration, the $T_2$ time is not limited by the $^{13}$C concentration, but by other effects \cite{mizuochi_coherence_2009} (e.g. longitudinal spin lattice relaxation, $T_1$).
For a diamond with natural abundance of $^{13}$C ($1.1\%$), a $T_2$ time of $0.65\,$ms was reported \cite{mizuochi_coherence_2009}, corresponding to a minimal hyperfine interaction of $\approx 1.5\,$kHz for a detectable $^{13}$C.
For performing high fidelity manipulation, we assume a minimal value of $\approx 5\,$kHz.
This corresponds to a maximal distance of $r_{\rm{max}}$ $\approx 1.5\,$nm.
Within this radius, there are $\frac{8}{(0.357 \rm{nm})^3}\frac{4}{3}\pi(1.5 \rm{nm})^3 \approx 2500$ diamond lattice positions (where $0.357\,$nm is the lattice constant of diamond).
For a $^{13}$C concentration of $1.1\%$, this yields, on average, 27 detectable $^{13}$C nuclear spins.

\section{Creation of a solid immersion lens}
\begin{figure}
\includegraphics[width=0.45\textwidth]{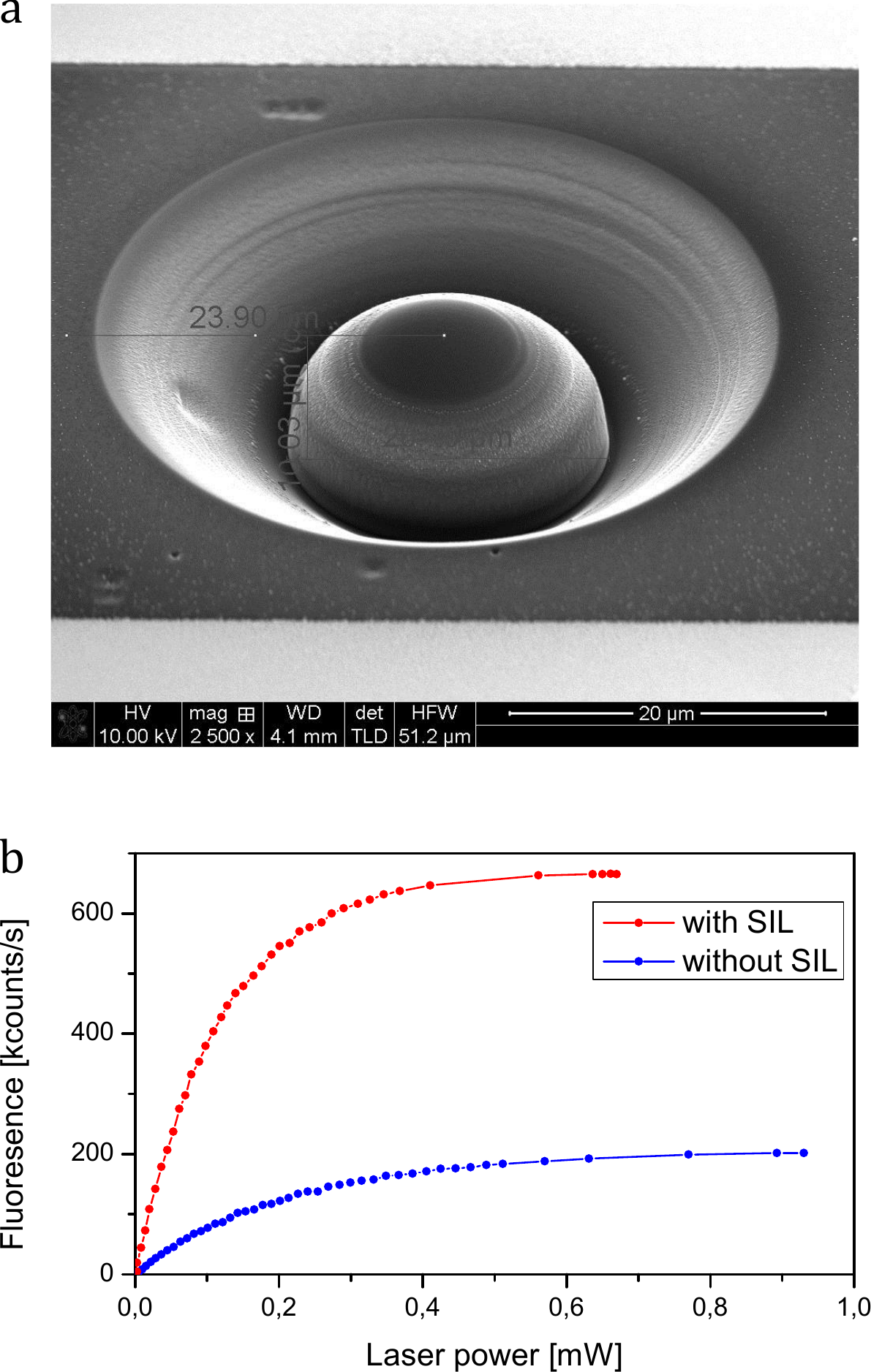}
	\caption{\textbf{a}, Image of the SIL in diamond.
	\textbf{b}, Saturation curves of the NV with and without the SIL.
	}
\label{fig:sil}
\end{figure}

All measurements of the state of the NV and surrounding nuclear spins are based on detection of the NV fluorescence.
Therefore, efficient photon detection is beneficial, as it increases measurement accuracy and reduces measurement time.
For single shot readout, measurement fidelity and hence state initialization fidelity is increased. 
However, much of the fluorescence of the NV is lost at the diamond surface due to total internal reflection.
This problem can be overcome by creating a solid immersion lens (SIL) \cite{marseglia_nanofabricated_2011}.
The SIL is a solid half sphere with the emitter (NV) at its central point, such that the fluorescence of the NV will hit the solid surface always perpendicular.
Here, we created a SIL directly in the diamond around the NV by focused ion beam milling, see Fig. \ref{fig:sil}a.
This increases the detected fluorescence of the NV by over a factor of 3 (Fig. \ref{fig:sil}b).


\section{Details of the measurement sequence}
The first step for every measurement is the initialization of nuclear spins.
This is achieved by measuring the state of the nuclear spin register, until it is in the desired state.
For technical reasons, we always perform the full measurement sequence, and finally only use the results for which the state was initialized.
Since the probability for successful initialization decreases with the number of states, we also use swap-like gates to transfer the initialized electron spin state to each nuclear spin, as shown in Fig. 2a of the manuscript.
Due to imperfect gates and non-perfect electron initialization, this increases the success probability of initialization of the whole register by a final single shot measurement to around 50\%.
Another important issue of the NV is its charge state.
Studies have shown that for typical measurement conditions, the NV is in the undesired neutral charge state with around 30 \% probability, which shows short coherence times \cite{waldherr_dark_2011, aslam_photo-induced_2013}.
Here, we use charge state postselection as demonstrated in \cite{waldherr_high_2011}.
To this end, first a controlled $\pi$-pulse is applied on the electron spin, only if the nuclear spins are in the initial state.
Next, we apply a controlled $\pi$-pulse on the $^{14}$N from $m_{I} =$ +1 to 0, depending on the electron spin state.
These two pulses will only work if the NV is properly initialized in its negative charge state.
Otherwise, the $^{14}$N will remain in the $m_{I} = +1$ state, which can be checked by a final single shot measurement.
After these two pi pulses, we start the actual experiment, as described in the manuscript, using the $^{14}$N $m_{I} = 0, -1$ subspace as a qubit.
Finally, we readout the state of the nuclear spins.
First, we measure if the $^{14}$N is still in the $m_{I} = +1$ state, in which case the NV was in the wrong charge state, and the result is discarded.
Subsequently, we measure the $^{14}$N $m_{I} = -1$ state and the states of the two $^{13}$C spins.

\begin{figure}
\includegraphics[width=0.45\textwidth]{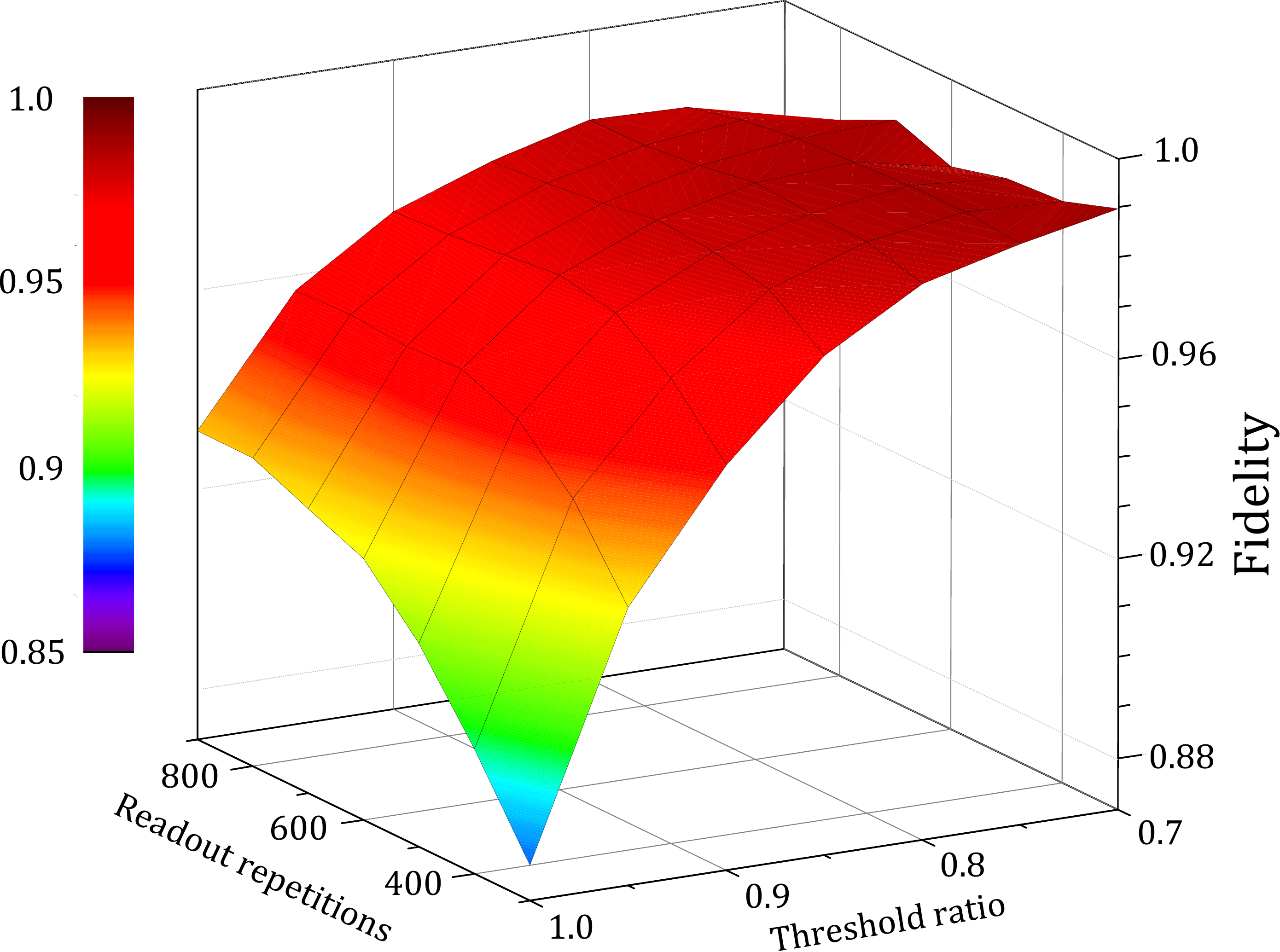}
	\caption{Dependence of initialization fidelity of the whole spin register on the number of readout repetitions and the relative shift of the initialization threshold.
	}
\label{fig:fidelity}
\end{figure}

For the initialization of the nuclear spin register, we can in principle reach near perfect initialization, by shifting the threshold in fig. 1e of the main manuscript further down.
Fig. \ref{fig:fidelity} shows the measured initialization fidelity depending on the duration of the single shot readout (in number of readout steps) and the relative threshold shift.
An initialization fidelity approaching $99\%$ can be achieved.
This method comes at the trade-off of increased measurement time, because the probability of a successful initialization will be reduced.

\section{Optimal control of electron spin}
\begin{figure*} 
\includegraphics[width=0.65\textwidth]{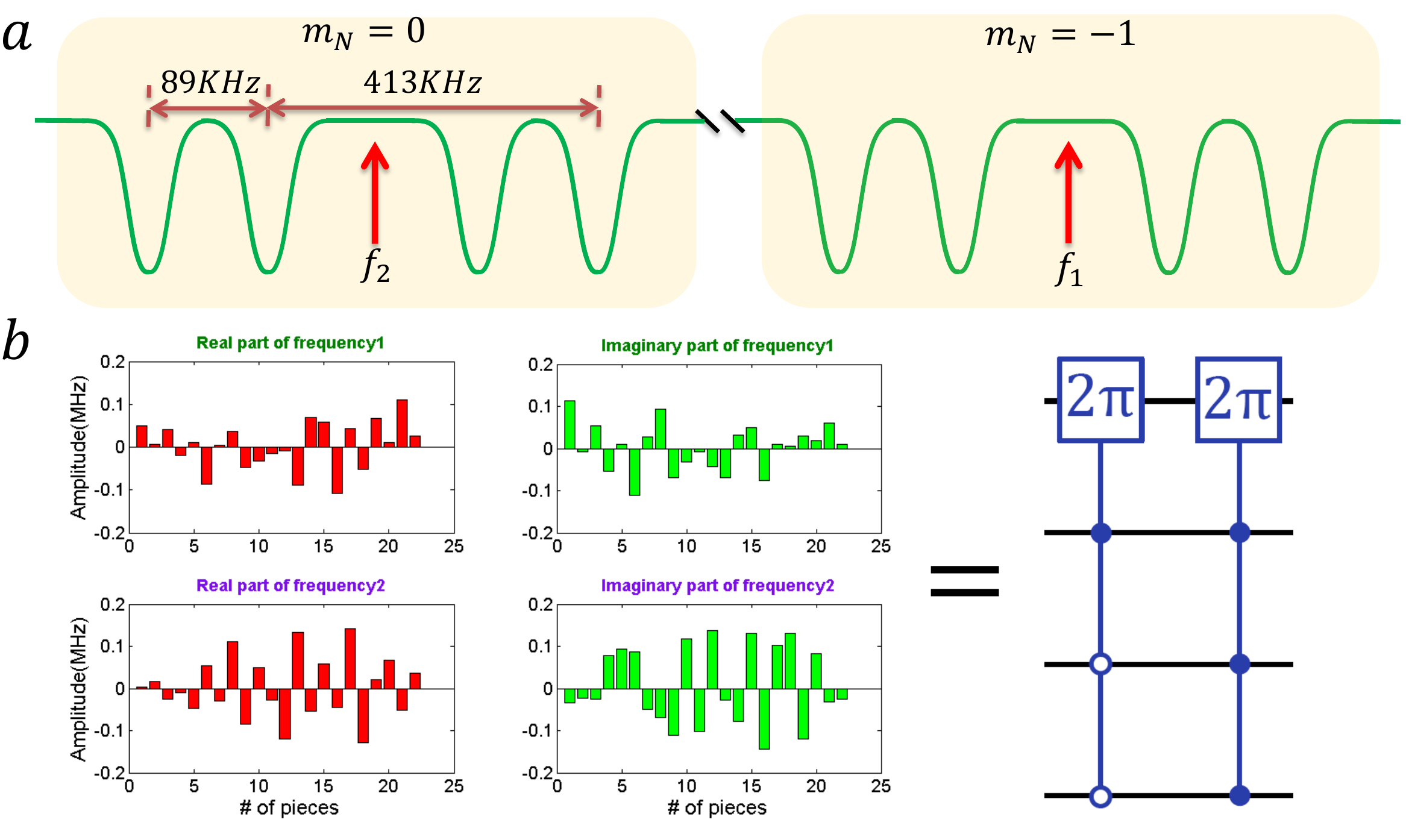}
	\caption{\textbf{a} The frequency of two MWs applied in the experiment.
	\textbf{b}, The pulse sequence on the left side is made by piece-wise constant control amplitudes, where each bar has a duration of 1.46 $\mu$s.
	   It realizes the controlled gate on the electron spin given on the right side.
	}
\label{fig:pulse}
\end{figure*}

Due to the weak coupling to the two $^{13}$C spins, selective rotations on specified transitions with standard rectangular microwave pulses will induce an additional unwanted operation on others being off-resonance. 
To suppress the off-resonance effect and achieve a high-fidelity control of the electron spin, we apply optimal control to the electron spin. 
Two MW frequencies ($f_1$,$f_2$) are imposed simultaneously to cover the whole spectrum and realize arbitrary control of the electron spin (Fig. \ref{fig:pulse}a). 
The amplitude and phase of each MW frequency are taken as free parameters to obtain the desired control. 
The pulse sequence is optimized using the GRAPE-algorithm \cite{khaneja_optimal_2005, machnes_comparing_2011}.
In the experiment, the optimal pulse sequence is realized by an arbitrary waveform generator (AWG; Agilent Technologies M8190A). 
Fig. \ref{fig:pulse}b shows an optimal pulse sequence applied to prepare a GHZ state of nuclear spins in the experiment. 
It is indeed a controlled phase (CPhase) gate on the electron spin, which maps the nuclear spin state $|100\rangle$ and $|111\rangle$ to state $-|100\rangle$ and $-|111\rangle$. 
The calculated fidelity is above $99\%$ even for small detuning ($\pm$20 KHz), which means that the pulse is robust in the presence of inhomogeneous line broadening of the electron spin. 
In the actual implementation, the instability of the MW amplitude is found to be one factor limiting the performance of the pulse.

\section{Mermin inequality}
A first test for the non-deterministic and non-local structure of quantum mechanics was proposed by Bell in 1964 for the case of two entangled particles \cite{bell_einstein_1964}.
A general inequality to measure this behaviour of quantum mechanics for multiple particles was derived by Mermin \cite{Mermin_extreme_1990}.
For three particles, the inequality is
\begin{equation}\label{eq:mermin}
\left|\left\langle\sigma_x\sigma_x\sigma_y\right\rangle+\left\langle\sigma_x\sigma_y\sigma_x\right\rangle+\left\langle\sigma_y\sigma_x\sigma_x\right\rangle-\left\langle\sigma_y\sigma_y\sigma_y\right\rangle\right| \leq 2
\end{equation}
for any deterministic, local theory.
However, the state
\begin{equation}\label{eq:mermin_state}
\left|\psi\right\rangle = \frac{1}{\sqrt{2}}(\left|000\right\rangle+i\left|111\right\rangle)
\end{equation}
yields the value 4 for (\ref{eq:mermin}).
In our case, the prepared state is
\begin{equation}
\left|\psi\right\rangle = \frac{1}{\sqrt{2}}(\left|0\overline{y}\overline{y}\right\rangle+i\left|1yy\right\rangle),
\end{equation}
where $\left|y\right\rangle = \left|0\right\rangle + i\left|1\right\rangle$ and $\left|\overline{y}\right\rangle = \left|0\right\rangle - i\left|1\right\rangle$.
This state can be derived from (\ref{eq:mermin_state}) by $\pi/2$ rotations around the x axis on qubit two and three.
Thereby, $\sigma_y$ in (\ref{eq:mermin}) becomes $\sigma_z$ for these two qubits, yielding the inequality
\begin{equation}\label{eq:mermin2}
\left|\left\langle\sigma_x\sigma_x\sigma_z\right\rangle+\left\langle\sigma_x\sigma_z\sigma_x\right\rangle+\left\langle\sigma_y\sigma_x\sigma_x\right\rangle-\left\langle\sigma_y\sigma_z\sigma_z\right\rangle\right| \leq 2.
\end{equation}
To measure the single terms, we transform each term to $\left\langle\mathbbm{1}\mathbbm{1}\sigma_z\right\rangle$.
This can be achieved by a unitary operation $U$ according to $\left\langle\psi\left|A\right|\psi\right\rangle = \left\langle\psi'\left|A'\right|\psi'\right\rangle$ for $\left|\psi'\right\rangle = U\left|\psi\right\rangle$ and $A' = UAU^{\dagger}$.
The operations are shown in table \ref{tab:process}.
\begin{table}[ht]
\centering
\begin{tabular}{|c |c |}
\hline
Measure & Operation \\[0.5ex]
\hline  \hline
$\left\langle\sigma_x\sigma_x\sigma_z\right\rangle$ & $\left(\frac{\pi}{2}\right)_{1, y}\left(\frac{\pi}{2}\right)_{2, y}\left(-\frac{\pi}{2}\right)_{3, y}\left(CPhase\right)\left(\frac{\pi}{2}\right)_{3, y}$ \\   \hline
$\left\langle\sigma_x\sigma_z\sigma_x\right\rangle$ & $\left(\frac{\pi}{2}\right)_{1, y}\left(CPhase\right)\left(\frac{\pi}{2}\right)_{3, y}$ \\   \hline
$\left\langle\sigma_y\sigma_x\sigma_x\right\rangle$ & $\left(\frac{\pi}{2}\right)_{1, x}\left(\frac{\pi}{2}\right)_{2, y}\left(CPhase\right)\left(\frac{\pi}{2}\right)_{3, y}$ \\   \hline     
$\left\langle\sigma_y\sigma_z\sigma_z\right\rangle$ & $\left(\frac{\pi}{2}\right)_{1, x}\left(-\frac{\pi}{2}\right)_{3, y}\left(CPhase\right)\left(\frac{\pi}{2}\right)_{3, y}$ \\   \hline
\end{tabular}
\caption{Measurement procedure for the process fidelity. \label{tab:process}}
\end{table}
The Cphase gate is described in the main manuscript; here it is conditional on the states $\left|011\right\rangle$ and $\left|101\right\rangle$.

\section{State tomography}
The density matrix of the entangled GHZ and W state is measured by state tomography.
Therefore, the density matrix $\rho$ is expanded as
\begin{equation}\label{eq:rho}
\rho = \sum_i a_i A_i,
\end{equation}
where the $A_i$ form a basis of mutually orthogonal operators.
In the case of a single qubit, we would use the Pauli matrices $\mathbbm{1}$, $\sigma_x$, $\sigma_y$, $\sigma_z$.
For multiple qubits, the $A_i$ are all combinations of the Pauli matrices for all qubits, i.e. $A_i = \left\{ \mathbbm{1}^{(1)}\mathbbm{1}^{(2)}...\mathbbm{1}^{(n)}, \mathbbm{1}^{(1)}\mathbbm{1}^{(2)}...\sigma_x^{(n)}, ...,  \sigma_z^{(1)}\sigma_z^{(2)}...\sigma_z^{(n)}\right\}$, where the superscripts indicates on which qubit the operator acts.
The $a_i$ in (\ref{eq:rho}) are given by the expectation value $\rm{Tr}\left( A_i\rho \right)$ of the corresponding operator.
These expectation values can be determined experimentally.
Since only the expectation value of the $\sigma_z$ operator can be measured directly, we apply $\pi/2$ rotations around the $y$ and $x$ axis of a spin to get the expectation values of $\sigma_x$ and $\sigma_y$.
Solving (\ref{eq:rho}) with the measured $a_i$ yields the state of the system.
Note that due to statistical measurement errors, the measured density matrix may not be positive semi-definite.

\section{Process tomography}
A general characterization of a quantum mechanical process $\mathcal{E}$ on a system is obtained by expanding the density matrix $\rho$ into a basis of orthonormal states $\left|\psi_i\right\rangle$, and measuring the effect of the operator $E_i$ of the process on each of these states:
\begin{eqnarray}
\rho &&\ = \sum_i p_i \left|\psi_i\right\rangle \left\langle \psi_i\right|, \nonumber\\
\mathcal{E}(\rho) &&\ = \sum_i E_i p_i \left|\psi_i\right\rangle \left\langle \psi_i\right| E_i^{\dagger} \label{eq:process}.
\end{eqnarray}
The $E_i$ can be measured by initializing the states $\left|\psi_i\right\rangle$ and performing state tomography after the process.
Furthermore, we can do the expansion
\begin{equation}
E_i = \sum_m \chi_{im} A_m,\label{eq:exp1}
\end{equation}
where the $A_m$ form a basis of mutually orthogonal operators.
Inserting (\ref{eq:exp1}) into (\ref{eq:process}) yields
\begin{equation}
\mathcal{E}(\rho) = \sum_{mn} \chi_{mn} A_m \rho A_n^{\dagger},
\end{equation}
where the $\chi_{mn} = \sum_i \chi_{im}\chi_{in}$ form the process matrix $\chi$.

For the error correction, we only consider the effect of the process on the qubit that is carrying the information after the correction.
Additionally, we are only interested in the fidelity $F$ of the measured process and the ideal process $\chi_{\rm{id}}$, which is
\begin{equation}
F = \rm{Tr}(\chi_{\rm{id}}\chi).
\end{equation}
Here, we chose the Pauli matrices $A_m = \left\{ \mathbbm{1}, \sigma_x, \sigma_y, \sigma_z \right\}$ as the operator basis.
The ideal process for the error correction is the identity, and therefore $\chi_{11}^{\rm{id}} = 1$, and all other values being zero.
The fidelity is then $\chi_{11}$.
Table \ref{tab:fidelity} shows the measurements we perform to obtain
\begin{equation}
\chi_{11} = \frac{1 + \left(r_{z,z}-r_{-z,z}+r_{x,x}-r_{-x,x}+r_{y,y}-r_{-y,y}\right)}{4},
\end{equation}
by noting that $\chi_{11}+\chi_{22}+\chi_{33}+\chi_{44} = 1$.

\begin{table}[ht]
\centering
\begin{tabular}{|c |c |c |}
\hline
Init & Measure & result \\[0.5ex]
\hline  \hline
$\left|z\right\rangle$ & $\sigma_z$ & $r_{z,z} = \chi_{11}+\chi_{14}+\chi_{41}+\chi_{44}$ \\   \hline
$\left|-z\right\rangle$ & $\sigma_z$ & $r_{-z,z} = -\chi_{11}+\chi_{14}+\chi_{41}-\chi_{44}$ \\   \hline
$\left|x\right\rangle$ & $\sigma_x$ & $r_{x,x} = \chi_{11}+\chi_{12}+\chi_{21}+\chi_{22}$ \\   \hline
$\left|-x\right\rangle$ & $\sigma_x$ & $r_{-x,x} = -\chi_{11}+\chi_{12}+\chi_{21}-\chi_{22}$ \\   \hline
$\left|y\right\rangle$ & $\sigma_y$ & $r_{y,y} = \chi_{11}+\chi_{13}+\chi_{31}+\chi_{33}$ \\   \hline
$\left|-y\right\rangle$ & $\sigma_y$ & $r_{-y,y} = -\chi_{11}+\chi_{13}+\chi_{31}-\chi_{33}$ \\   \hline
\end{tabular}
\caption{Measurement procedure and theoretical results for the process fidelity. \label{tab:fidelity}}
\end{table}
